# Graphitic $C_3N_4$ Sensitized $TiO_2$ Nanotube Layers:

# A Visible Light Activated Efficient Antimicrobial Platform

Jingwen Xu,[a] Yan Li,[a] Xuemei Zhou,[b] Yuzhen Li,[a] Zhi-Da Gao,[a] Yan-Yan Song[a] and Patrik Schmuki[bc]

a. College of Sciences, Northeastern University, Box 332, Shenyang 110004, China. E-mail: yysong@mail.neu.edu.cn. Tel.: +86-24-83687659.

b. Department of Materials Science, WW4-LKO, University of Erlangen-Nuremberg, Martensstrasse 7, D-91058 Erlangen, Germany. E-mail: schmuki@ww.uni-erlangen.de. Tel.: +49-9131-852-7575.

c. Department of Chemistry, Faculty of Science, King Abdulaziz University, P.O. Box 80203, Jeddah 21569, Saudi Arabia

**Abstract:**

In this work, we introduce a facile procedure to graft a thin graphitic $C_3N_4$ (g-$C_3N_4$) layer on aligned $TiO_2$ nanotube arrays (TiNT) by one-step chemical vapor deposition (CVD) approach. This provides a platform to enhance the visible-light response of $TiO_2$ nanotubes for antimicrobial applications. The formed g-$C_3N_4$/TiNT binary nanocomposite exhibits excellent bactericidal efficiency against E. coli as a visible-light activated antibacterial coating.

Sterile or bactericidal surfaces are of a wide interest in healthcare related fields such as hospital environments.[1] There are several strategies to create a bacteria-free surface such as using disinfectant chemical treatments by alcohol, formaldehyde, or hydrogen peroxide,[2] or engineering antimicrobial coatings that contain bactericidal elements or compounds such as Ag, Cu, iodine, fluorine, phenol or heavy metals.[3] However, many of these approaches provide only a time-limited antimicrobial activity,[4] resistance to the anti-infectious agents,[5] and can release toxic elements or compounds to environment.[6]

Another conceptually different strategy is to trigger bacterial inactivation by employing a photocatalytically active coating.[7] Generally a number of photocatalysts can induce bacterial inactivation by the generation of highly reactive species (e.g. hydroxyl radicals and superoxides) or by direct valence band hole ejection that cause fatal damage to microorganisms.[8] During the photocatalysis process, typically a decomposition of the cell membrane occurs, which can directly lead to the leakage of minerals, proteins, and genetic material, and thus results in cell death.[9]

As one of the most important photocatalysts, titanium dioxide ($TiO_2$) has attracted considerable attention in the antibacterial field, since its antimicrobial property was first reported in 1985.[10] Titanium and its alloys are widely used in advanced medical applications such as orthopedics, dentistry, as well as components in medical devices and instruments. Thus the surface of such devices coated with a native or an anodic $TiO_2$ layer can show an intrinsic photocatalytic activity. Unfortunately, due to the wide band-gap of $TiO_2$ (~3.2 eV for anatase), the related photocatalytic properties (i.e. bacterial inactivation, self-cleaning and self-sterilization) need UV light to be triggered. $TiO_2$ is thus of a limited value, particularly in indoor environments (with low UV levels), such as in a hospital environment where preventing attachment and subsequent colonization of bacteria onto surfaces would be highly desired.[11] Therefore, strategies to activate a visible-light photocatalytic response have become one of the most important issues in developing $TiO_2$ based antimicrobial coatings.[12]

Very recently, graphitic carbon nitride (g-$C_3N_4$), a polymeric semiconductor, has attracted considerable interest due to a band-gap of ~2.69 eV and its adequate band-positions for various red-ox reactions.[13] However, its photocatalytic activity is unsatisfactory due to the low quantum efficiency.[14] To resolve this problem, many attempts have been carried out to improve its photocatalytic performance, such as non-metal doping,[15] preparation of nano/porous g-$C_3N_4$,[16] and formation of band-gap matched heterojunction between g-$C_3N_4$ and other photocatalysts such as $TiO_2$, ZnO, and $BiPO_4$.[17] Most of these studies have focused on the performance in the photocatalytic decomposition of organic pollutants and energy conversion.[18] No work reported so far has dealt with the photocatalytic inactivation of bacteria under visible-light irradiation using g-$C_3N_4$ sensitized $TiO_2$ or other wide band-gap semiconductors or its use in an antimicrobial coating.

In the present work, we designed a one-step chemical vapor deposition (CVD) approach to deposit a small amount of g-$C_3N_4$ on aligned $TiO_2$ nanotube layers. Previous work showed that such nanotubular or nanoporous $TiO_2$ layers can be grown on a broad range of titanium geometries as well as on titanium alloys by a self-organizing electrochemical anodization process.[19] These nanotubular or nanoporous structures provide highly defined nanoscale compartments that can be used as efficient payload scaffolds.[20] The g-$C_3N_4$ layer used here can be synthesized from a simple precursor via a polycondenzation reaction (as outlined in the SI). We then studied the photocatalytic antibacterial property of the g-$C_3N_4$/TiNT nanocomposites under visible-light irradiation by inactiviting of *Escherichia coli* (*E. coli*, a Gram-negative bacterium). The concept being illustrated in Fig. 1a relies on the photocatalytic generation of highly oxidative species that deactivate bacteria cells under visible light. The investigated approach offers a facile and cheap way to produce antibacterial surfaces and can be applied to a broad range of material surfaces. Thus it represents a promising coating for reducing infections (e.g. nosocomial infection in hospitals) and other indoor applications where only visible light is available.

Fig. 1b shows the SEM images of the TiNT layer grown on the Ti substrate. The aligned nanotube layers used in this work, with ~2.0 μm thickness and an average tube diameter of ~150 nm (Fig. 1b and 1c), are formed under optimized electrochemical anodization conditions (as described in the Supporting Information, ESI*). In order to obtain an optimized photocatalytic activity, the amorphous $TiO_2$ nanotubes were converted to anatase by annealing in air. These tube layers were then decorated with g-$C_3N_4$ (as described in the SI). The SEM images in Fig. 1d and 1e clearly show that after the CVD decoration process the surface of the $TiO_2$ nanotubes is covered by patches of a g-$C_3N_4$ nanofilm. After deposition the colour of the samples changes from brown to black (see inset optical images in Fig. 1c and 1e).

These g-$C_3N_4$ loaded layers (in contrast to plain $TiO_2$ NT layers) show a visible light photo response as evident from photocurrent spectra. Fig. 1f shows photocurrent versus wavelength data recorded in a 0.1 M $Na_2SO_4$ aqueous solution (acquired as described in the SI) for nanotube layers loaded with different amounts of g-$C_3N_4$. The non-decorated tubes only show a UV response that corresponds to the band-gap of anatase of 3.2 eV. For the g-$C_3N_4$ loaded samples, in the visible-light region, the samples exhibit considerably enhanced photocurrents with an onset at ~2.4 eV. The highest visible response is obtained from a loading of 15 mg precursor. Therefore this loading was selected for further analysis and application. As shown in the X-ray diffraction (XRD) patterns of Fig. 1g, the appearance of a peak at 27.5° in g-$C_3N_4$/TiNT sample, which is corresponding to the interlayer stacking of aromatic segments, further confirms the formation of g-$C_3N_4$ on TiNT.

In Fig. 2a, the X-ray photoelectron spectroscopy (XPS) survey scan illustrates that the as-prepared sample contains elements of Ti, O, C and N. Fig. 2b shows the N 1s high-resolution spectrum, which is mainly composed of two peaks centered at 398.4 and 400.1 eV corresponding to the $sp^2$-bonded N involved in the triazine rings (C-N=C) and the bridging N atoms in N-$(C)_3$, correspondingly.[16] The small peak located at 403.7 eV is ascribed to the amino functions carrying hydrogen (N-H). Fig. 2c shows three peaks located at 284.8, 286.6, and 287.9 eV in the C 1s spectra. The peak at 284.6 eV is ascribed to $sp^2$ C=C bonds, the peak at 286.6 can be assigned to the $sp^2$ hybridized carbon atoms bonded to three nitrogen atoms in the g-$C_3N_4$ layer, and the peak at 287.9 eV is related to the $sp^2$ carbon atoms in the aromatic ring. The signals of O 1s (Fig. S1a, ESI*) and Ti 2p (Fig. S1b, ESI*) are typical of $TiO_2$ nanotubes. Additionally, FT-IR spectra of g-$C_3N_4$/TiNTs were also recorded in Fig. 2d. The broad bands in the 3000~3700 $cm^{-1}$ region are attributed to adsorbed $H_2O$ molecules and N-H vibration modes from the uncondensed amine groups. The bands in the range of 1200~1600 $cm^{-1}$ relate to the stretching mode of C–N heterocycles. The sharp band observed at 804 $cm^{-1}$ is corresponding to the breathing mode of triazine units. A typical band of $TiO_2$ can be seen at ~470 cm$^{-1}$. These results confirm the successful synthesis and loading of g-$C_3N_4$ on $TiO_2$ nanotubes.

Concerning the prospective visible photocatalytic application of g-$C_3N_4$/TiNT samples, the photo-induced inactivation of bacteria was investigated. Escherichia coli (*E. coli*), a typical bacterium responsible for many infections in daily life, is used as a model in the present work. The photocatalytic antibacterial activity of g-$C_3N_4$/TiNT under visible light irradiation is evaluated in terms of the survival ratio of *E. coli*, which is defined as the ratio of the number of bacteria colonies present on the test surface and the number of colonies on a reference glass substrate (as glass substrates exhibit hardly any antibacterial activity). In Fig. 3a, the bare anatase and amorphous samples only show poor antibacterial activities with a survival ratio of ~72-74%. Noticeably, the g-$C_3N_4$/TiNT samples present a significantly improved antibacterial activity (the survival ratio of bacteria is only ~16%) compared to the g-$C_3N_4$ decorated glass slide (g-$C_3N_4$/glass shows a survival ratio of ~86% when irradiated by visible light). In addition, no obvious inactivation was observed in the dark control (catalysts without light irradiation) experiment, suggesting the nontoxicity of as-prepared g-$C_3N_4$/TiNT nanocomposite to the *E. coli* cells during the test period. These results suggest that the photoinduced active species are the key for the bacterial inactivation on the g-$C_3N_4$/TiNT antimicrobial coating, and the formation of g-$C_3N_4$/$TiO_2$ can promote the antibacterial efficiency in visible light.

To correlate the antibacterial efficiency with the photocatalytic activities of g-$C_3N_4$/TiNT layers, the survival ratios were measured on a series of g-$C_3N_4$/TiNT samples, which were prepared using different amounts of precursor. Fig. **3b** shows the bacterial survival ratios as a dependence on the precursor loading (and for reference the visible light photocurrent response of these samples at 420 nm **Fig. 3b**). The results suggest that the bactericidal efficiency is directly related to the photocatalytic activity in visible light. The best bactericidal efficiency is achieved for the sample that reaches the highest visible photocurrent. In other words, these bactericidal results show that a loading of 15 mg of melamine is the optimal precursor amount for preparing highest efficiency g-$C_3N_4$/TiNT layers.

The photocatalytic antibacterial activity of g-$C_3N_4$/TiNTs was further studied for several bactericidal cycles (Fig. 3c). Some decrease of antibacterial activity in repeated cycles can be observed for the g-$C_3N_4$/TiNTs samples − this may particularly originate from photocorrosion of the g-$C_3N_4$ film under light irradiation for long periods as well as an accumulating coverage of the surface with dead bacterial cells (some dead bacterial cells remain attached to the coating surface even after washing the surface by ethanol and DI water thoroughly after each bactericidal cycle). In addition, the storage stability of g-$C_3N_4$/TiNTs samples is very high – the samples keep ~99% and 94% of their bactericidal activity against *E. coli* after being stored in dark and visible light ($\lambda$>450 nm, 50 $\mu$Wcm$^{-2}$) at room temperature for 3 months, respectively.

To illustrate the effect of g-$C_3N_4$/TiNT antibacterial coatings on the morphology of *E. coli*, we took SEM images before and after photo-inactivation (Fig 3d-e). Before illumination, *E. coli* cells present an intact cell structure and a well-defined rod shape. After the photocatalytic inactivation under visible-light irradiation (Fig. 3e), the outer membrane of bacteria cells exhibits severe damage.[21] The optical images represent the density of survived *E. Coli* on the glass slide (Fig. 3d inset) and on g-$C_3N_4$/TiNT (Fig. 3e inset) after photocatalytic inactivation (the image are taken following incubation on solid LB medium for 24 h). Clearly, only a small percentage of bacteria colonies survive after the visible-light induced photocatalytic treatment on g-$C_3N_4$/TiNT. This result further illustrates the efficiency of inactivation of *E. coli* under visible-light irradiation using g-$C_3N_4$/TiNT coatings.

The assumed underlying mechanism for the anti-bacterial activity of the g-$C_3N_4$/$TiO_2$ system is outlined in Fig. 1a. It is based on a classic photocatalytic reaction scheme where the key is the creation of photo-promoted electrons that react with environmental $O_2$ to form superoxide radicals and related strong active oxidizers. In the present case g-$C_3N_4$ acts as visible light absorber (sensitizer, Fig. S2, ESI*) that (based on relative energetic positions of the band) forms a junction with $TiO_2$ that aids electron hole separation, and thus suppresses recombination. Electrons excited to either the g-$C_3N_4$ or the $TiO_2$ conduction band are energetically in a position that can react with environmental $O_2$ to form superoxide species (Fig. S3a, ESI*). In other words, the presence of oxygen in such a scheme is the key for reactivity. And indeed we find that the antibacterial activity of g-$C_3N_4$/TiNT decreases significantly (the bacterial survival ratio increased to ~ 60%) if the photocatalytic bacterial inactivation experiments are performed in anaerobic conditions (as an anaerobic bacteria, *E. coli* can live without oxygen).

Moreover, superoxide can react to OH• radicals. In order to detect the presence of OH•, we employed a classic fluorescence approach based on the reaction of OH• with terephthalic acid (TA) to form 2-hydroxyterephthalic acid (which shows a characteristic blue fluorescence at 426 nm) (Fig. S3b, ESI*).[22] Under visible light illumination no obvious fluorescence can be detected for amorphous or anatase sample without g-$C_3N_4$ sensitizing. In contrast, for g-$C_3N_4$ hydroxyl radicals can be detected. Overall, these findings support the valence band mechanism outlined in **Fig. 1a**.

**Conclusions**

By combining $TiO_2$ nanotube layers and g-$C_3N_4$ nanofilms we have developed a binary nanocomposite which shows a significant bactericidal effect under visible light. The effect is based on the visible light induced photocatalytic formation of highly reactive oxidative species. These active species severely attack microorganisms, in our case *E. Coli*. However, we believe that the platform reported here can be used in a much wider range of visible light activated photocatalytic processes and applications.

**Acknowledgements**

This work was supported by the National Natural Science Foundation of China (No. 21322504, 21275026), the Fundamental Research Funds for the Central Universities (N140505001, N140504006), and the Program for Liaoning Excellent Talents in University (LJQ2013028). We also thank ERC, DFG and DFG-funcos for their support.

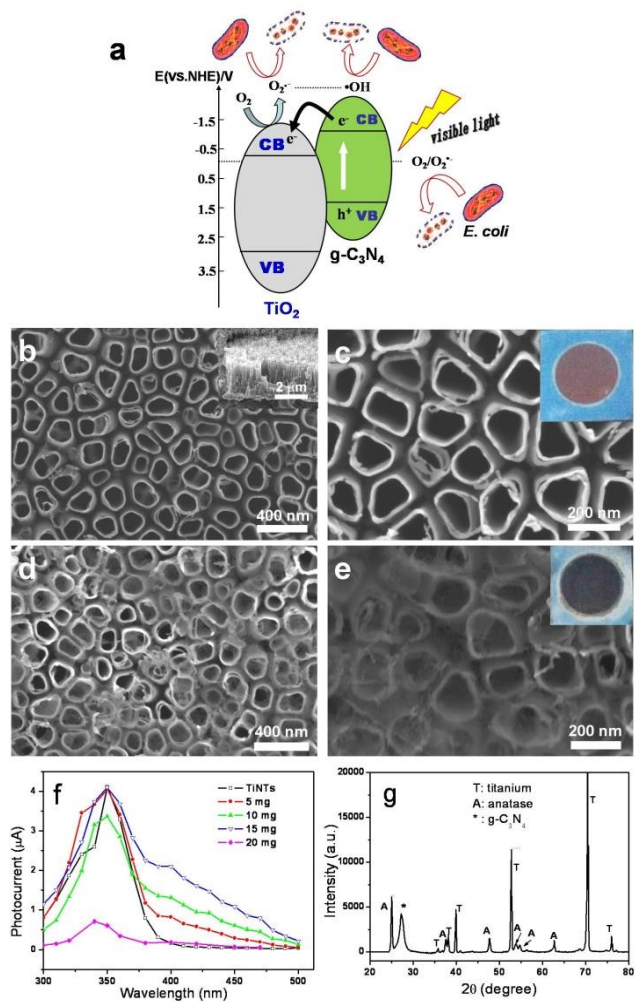

Fig. 1 (a) Scheme for the photocatalysis induced antibacterial mechanism of g-$C_3N_4$/TiNT under visible-light illumination. SEM images: top view of TiNT layer (b and c) and g-$C_3N_4$/TiNT layer (d and e). Insert: cross-sectional view of bare $TiO_2$ nanotube arrays and corresponding optical images for TiNT layer and g-$C_3N_4$/TiNT layer. (f) Photocurrent-wavelength responses of TiNTs and g-$C_3N_4$/TiNTs prepared from different amount of melamine as a precursor at a bias of +0.5 V (*vs* SCE) in 0.1 M $Na_2SO_4$. (g) XRD patterns of g-$C_3N_4$/TiNT layers prepared at 550 °C.

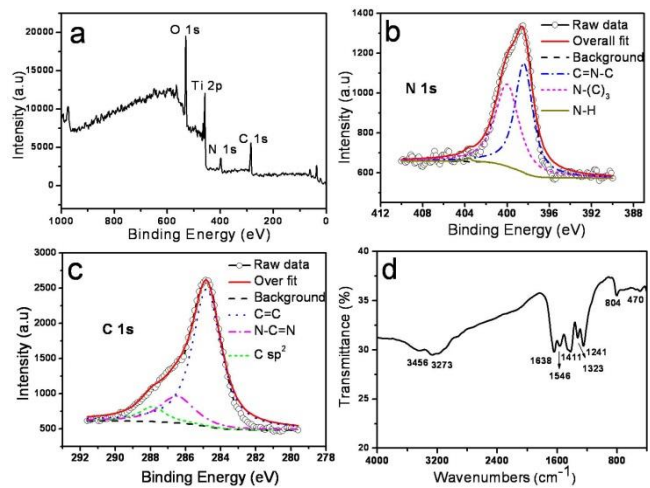

**Fig. 2** (a) XPS survey spectrum of g-$C_3N_4$/TiNTs, (b) N 1s and (c) C 1s for g-$C_3N_4$/TiNT, and (d) FT-IR spectrum of g-$C_3N_4$/TiNT.

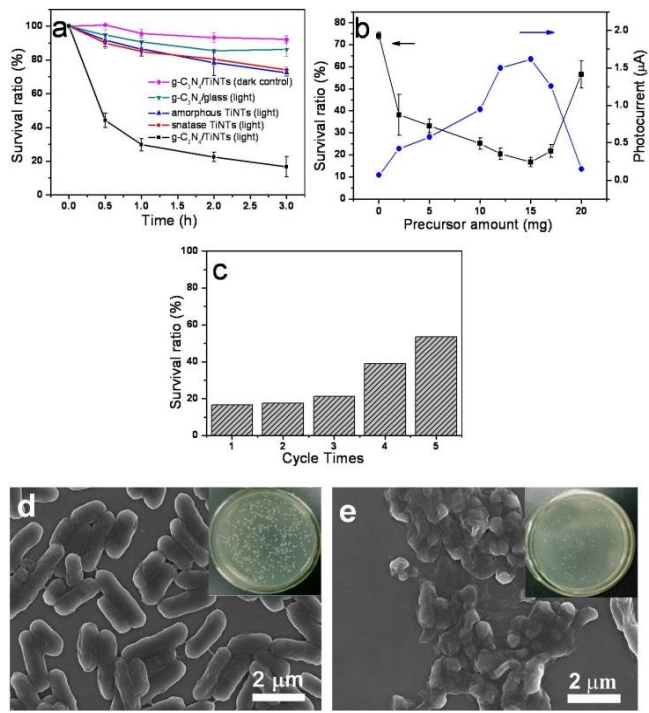

Fig. 3 (a) Survival ratio of *E. coli* on g-$C_3N_4$/glass, amorphous TiNT, anatase TiNT, and g-$C_3N_4$/TiNT layers by irradiating under visible-light, and the survival ratio of *E. coli* on g-$C_3N_4$/TiNT layers in dark. (b) The influence of dosage of melamine on the photocurrent (at 420 nm) of as-formed g-$C_3N_4$/TiNT samples and the survival ratio of *E. coli* after irradiated by visible light for 3 h. (c) The survival ratio *E. coli* on optimized g-$C_3N_4$/TiNTs sample under visible-light irradiation for five cycles. SEM images of *E. coli* without (d) and with (e) photoinduced inactivation on g-$C_3N_4$/TiNT surface by visible-light irradiation. The corresponding optical images of survival *E. coli* after incubation on solid LB medium at 37 °C for 24 h: (d inset) without and (e inset) with photoinduced inactive treatment on g-$C_3N_4$/TiNT surface by visible-light irradiation.

# Graphitic $C_3N_4$ Sensitized $TiO_2$ Nanotube Layers: A Visible Light Activated Efficient Antimicrobial Platform

Jingwen Xu,[a] Yan Li,[a] Xuemei Zhou,[b] Yuzhen Li,[a] Zhi-Da Gao,[a] Yan-Yan Song,[a]* Patrik Schmuki [b]*

[a] College of Sciences, Northeastern University, Box 332, Shenyang 110004, China

[b] Department of Materials Science, WW4-LKO, University of Erlangen-Nuremberg, Martensstrasse 7, D-91058, Erlangen, Germany

**Experimental Section**

**Preparation of TiO$_2$ nanotube arrays**

Ti foils (99.6% purity) were cleaned in ethanol and deionized water (DI) by sonication, and then dried with a nitrogen stream. The cleaned Ti foils were pressed together with a Cu plate against an O-ring sealed opening in an electrochemical cell (with 0.5 cm$^2$ exposed to the electrolyte) and then anodized in a glycerol/water (50:50) electrolyte with 0.27 M NH$_4$F at 30 V for 4 h at room temperature. Ti foils served as the working electrode, and a platinum sheet was used as the counter electrode. The as-formed TiNT layers were annealed in air at 550 $^{\circ}$C for 1 h to crystallize the TiO$_2$ to anatase before further decoration.

**Preparation of g-C$_3$N$_4$/TiNT samples**

g-C$_3$N$_4$ decoration was performed by a chemical vapor deposition (CVD) approach using melamine as a precursor. Specifically, a defined amount of melamine was added into a combustion boat with the TiNT layer top-down placed above the melamine powders (distance ≈ 1.0 cm). The system was sealed and heated at 550 $^{\circ}$C for 3 h to decorate the TiNT with the g-C$_3$N$_4$ polymer, and the as-prepared g-C$_3$N$_4$/TiNT samples were then stored in dark before experiment. The g-C$_3$N$_4$/glass samples were prepared in a similar way, but replacing the TiNT substrate with a glass slide (with a roughened surface by grinding) of the same surface area as the TiNT samples.

**Antibacterial test**

Gram-negative *Escherichia coli (E. coli)* DH5□ were selected as the model target microorganism for antibacterial tests. Before microbiological experiments, all glass ware was sterilized by autoclaving at 121 °C for 20 min. A 300 W xenon lamp with a UV cutoff filter (λ<420 nm) was used as light source for photocatalysis induced antibacterial study. For antibacterial experiments, 100 μL mixture of liquid Luria broth (LB) substrate containing 0.5% agar and 10$^5$ colony-forming unites per milliliter (CFU mL$^{-1}$) of *E. coli* was dropped onto the surface of g-C$_3$N$_4$/TiNT or TiNT samples. The light source was located 20 cm from the samples (50 mW cm$^{-2}$ illumination intensity). To quantify the antimicrobial results, the semi-solid bacteria solution was withdrawn after the photoinactivation step and then diluted serially with sterilized Milli-Q water to adjust the bacterial concentration (the aliquots were diluted to 10$^{-4}$ times in our present study) to ensure the growing bacterial colonies were legible. Then 100 μL of the treated solution was spread on solid LB medium. The colonies were continued to incubate at constant temperature of 37 °C for 24 h, and then counted to determine the survival bacteria numbers and bactericidal efficiency. For better comparison, bactericidal efficiency of amorphous and anatase TiNT layers was also studied by the same procedure. The survival ratio was determined by comparing the survival colony counts with the corresponding colony counts of control sample (sterilized glass sheet). All of the photocatalytic antibacterial experiments were repeated three times to give an average value.

**Trapping experiments for radicals and holes**

For the trapping of radicals and holes, 4 mL RhB (2 mg L$^{-1}$) solution was photodegraded by g-C$_3$N$_4$/TiNT layer under visible light (a 300 W xenon lamp with a UV cutoff filter ≥ 420 nm). The trapping experiments were carried out by adding benzoquinone (1.0 mmol L$^{-1}$, superoxide radical scavenger), ethylene diamine tetraacetic acid (10 mmol L$^{-1}$, hole scavenger), and methanol (1/15 volume ratio, hydroxyl radical scavenger) separately to get rid of radicals and holes during the photocatalysis process under visible-light irradiation. UV spectra were used to detect the degradation of RhB with time [S1-S4].

The generated hydroxyl radicals were detected by fluorescence (FL) spectra by using terephthalic acid (TA) as a probe molecule. The g-C$_3$N$_4$/TiNT sample was immersed into a 4 mL aqueous solution with terephthalic acid (3 mM), NaCl (0.1 M) and NaOH (0.01 M), and the FL spectra of the solution were measured for every 10 min with excitation wavelength at 315 nm [S5-6].

**Apparatus**

The morphologies of the g-C$_3$N$_4$/TiNT samples and bacteria were characterized using a field-emission scanning electron microscope (Hitachi FE-SEM S4800). X-ray photoelectron spectra (XPS) were recorded on a Perkin–Elmer Physical Electronics 5600 spectrometer. All the peaks are shifted with a standard of C 1s at 284.8 eV. The UV-vis absorption spectra were measured on a spectrophotometer (Perkin–Elmer, Lambda XLS+, USA). The photocurrents were recorded by measuring at 420 nm in an aqueous solution containing 0.1 M Na$_2$SO$_4$ with an applied bias of +0.5 V (*vs* saturated calomel electrode (SCE)).

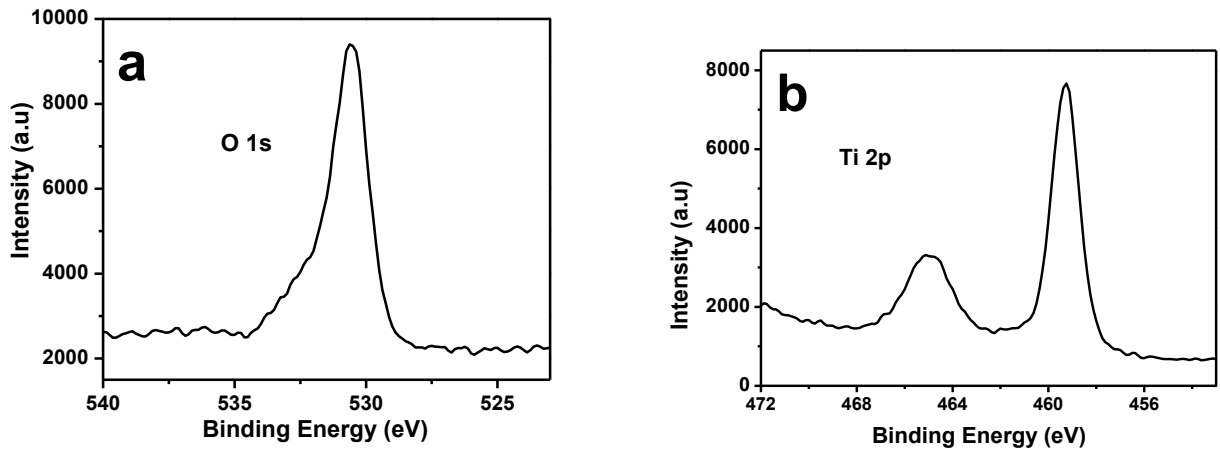

**Fig. S1.** XPS (a) O 1s and (b) Ti 2p spectra of g-$C_3N_4$/TiNTs layer.

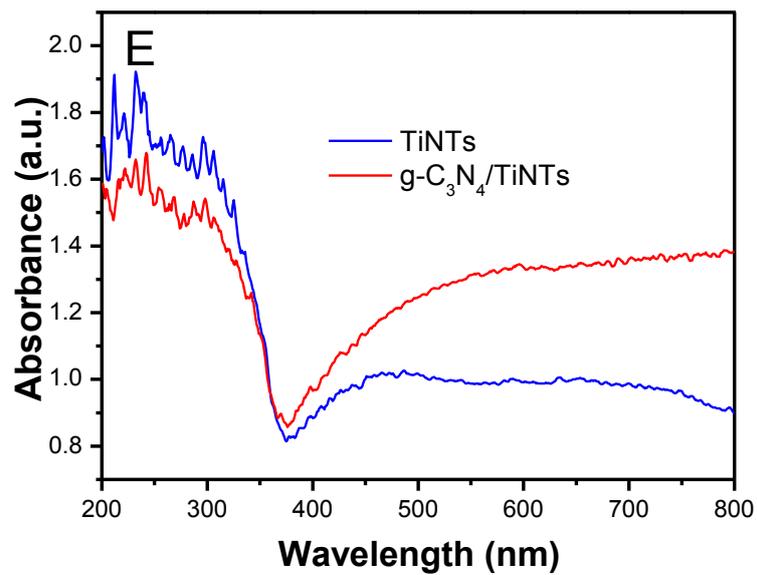

**Fig. S2.** UV-visible diffuse reflectance spectra of TiNT and g-$C_3N_4$/TiNT.

To evaluate visible-light activity introduced by the g-$C_3N_4$ decoration, we acquired reflectivity data. As shown in Fig. 2D, the decoration of g-$C_3N_4$ results in an obviously enhanced absorption in visible range.

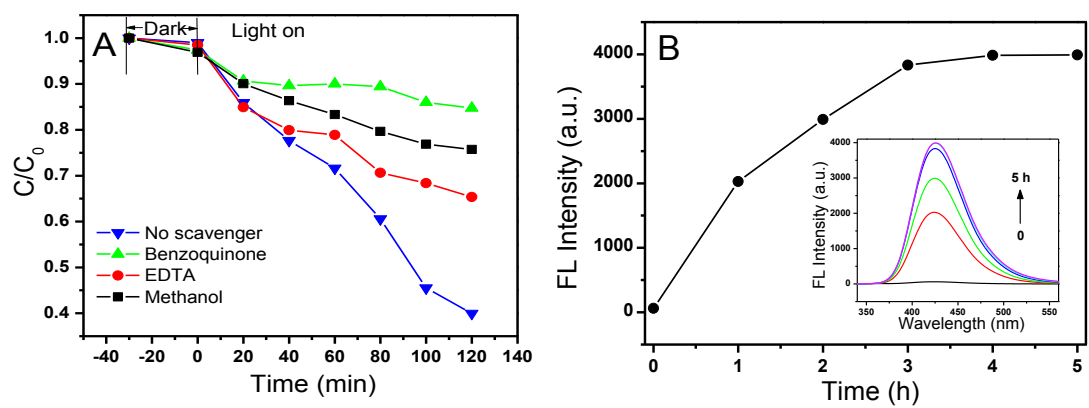

**Fig. S3** (A) Influence of different scavengers on the photocatalytic degradation of RhB by g-$C_3N_4$/TiNT samples under the irradiation of visible light. (B) Fluorescence intensity at 426 nm measured from the supernatant solution of terephthalic acid after irradiate g-$C_3N_4$/TiNT by visible light for different time. Insert: The time dependence fluorescence spectra.

In Fig. S3a, the photodegradation of organic dye (2 mg $L^{-1}$ RhB) is performed under visible light to further elucidate the possible reactive species that are directly involved in the damage of bacterial cells or organic molecules during the photocatalysis process. A significant decline in the photodegradation efficiency is observed when methanol (the scavenger agent for hydroxyl radicals), EDTA (the scavenger agent for holes) and benzoquinone (the scavenger agent for superoxide radicals) are separately added into the RhB solution. These results demonstrate that all the reactive species take part in the photocatalysis process, the superoxide radicals having the strongest influence [S1-S4]. Fig. S3b shows the increased fluorescence signals with irradiation time when g-$C_3N_4$/TiNT layer is illuminated by visible light in a TA containing solution. This result demonstrates the formation of hydroxyl radicals during the photocatalytic process [S5-6].